\documentstyle[times,pramana,epsf,floats]{ias}
\begin{document}
\mark{{Statistics of Resonances...}{Joshua Feinberg}}
\title{Statistics of Resonances in One Dimensional Continuous Systems}

\author{Joshua Feinberg}
\address{Physics Department, University of Haifa at Oranim, Tivon 36006, Israel\\ and\\ 
Physics Department, Technion, Haifa 32000, Israel}
\keywords{resonances, spectral determinant, disordered systems, Fokker-Planck equation, average density of resonances}
\pacs{03.65.Yz, 03.65.Nk, 72.15.Rn}
\abstract{
We study the average density of resonances (DOR) of a disordered one-dimensional continuous open system.  
The disordered system is semi-infinite, with white-noise random potential,  and it is coupled to the external world 
by a semi-infinite continuous perfect lead. Our main result is an integral representation for the DOR which involves 
the probability density function of the logarithmic derivative of the wave function at the contact point.} 
\maketitle
\section{Introduction}
Open systems typically give rise to resonances. A resonance is a long-living quasi-stationary state, which eventually decays into the continuum. 
Physically, it may be thought of as a particle, initially trapped inside the system, which eventually escapes to infinity. 

One common approach to studying resonances is based on the analytic properties of the scattering matrix $S({\cal E})$ in the complex energy plane. 
Resonances correspond to poles ${\cal E}_n = E_n - \frac{i}{2}\Gamma_n$ of $S({\cal E})$ on the non-physical sheet\cite{LL,BPZ}. 
In an alternative equivalent approach, one solves the Schr\"odinger equation subjected to the boundary condition of purely outgoing wave outside the range of the potential. 
This boundary condition, which describes a process in which a particle is ejected from the system,  renders the problem non-Hermitian. The Schr\"odinger equation with these boundary conditions leads to complex eigenvalues ${\cal E}_n$ which correspond to resonances \cite{LL,BPZ}. For a recent lucid discussion of resonances in quantum systems, with particular emphasis on the latter approach, see \cite{hatano,hf}.  

The outgoing-wave approach leads, in a natural way, to non-Hermitian effective hamiltonians ,whose complex eigenvalues are the resonances of the studied system \cite{wied,rotter,fyod-som}.  Such effective hamiltonians are very useful for studying resonances in scattering theory, including scattering in chaotic and disordered systems\cite{datta,kottos,ks1,ks2}. 

There are many examples of resonances in atomic and nuclear physics. Recently, there has been considerable interest in resonances which arise in chaotic and disordered systems. See \cite{kottos} for a recent review.  One of the main goals in these studies is computation of the distribution $P(\Gamma)$ of resonance widths. 
There is ample amount of work on computing $P(\Gamma)$ in one-dimensional disordered chains\cite{ks1,ks2,terraneo,texier,pinheiro,tit-fyod,weiss}. 
Numerical results presented in some of these works indicate that $P(\Gamma)\sim \Gamma^{-\gamma}$ in a large range of values of $\Gamma$, where the exponent
$\gamma$ is very close to $1$.

The present work was motivated by \cite{ks1,ks2}. In particular, an analytical approach was developed in \cite{ks2} for studying resonances, which is based on counting poles of the resolvent of the non-Hermitian tight-binding effective hamiltonian of the open chain. In the case of a semi-infinite disordered chain, coupled to a semi-infinite perfect lead, these authors have derived an exact integral representation for the density of resonances (DOR), valid for arbitrary disorder and chain-lead coupling strength. In the limit of weak chain-lead coupling (in which resonances are typically narrow) they were able to rigorously derive a universal scaling formula for the DOR, valid for any degree of disorder and everywhere inside  the unperturbed energy band of the closed chain. The $1/\Gamma$ behavior of the DOR follows from that formula. 

In this paper we shall lay out an analytical approach for computing the DOR of  a semi-infinite disordered system in the continuum. Our method is based on counting zeros of the spectral determinant of the continuous effective non-Hermitian hamiltonian.

\section{Resonances in the Semi-Infinite Disordered System and the Resonance Spectral Condition}
Let us consider a disordered system which occupies the domain $0\leq  x \leq L$. The system is assumed to be closed at $x=L$ (eventually, we shall take the limit $L\rightarrow\infty$). It is described by the Schr\"odinger hamiltonian\footnote{We use units in which $m=\hbar=1$.}
\begin{equation}\label{hamiltonian}
H_{\rm system~plus~contact} = -\frac{1}{2} \partial_x^2 + V(x) - \frac{\lambda}{2}\delta (x)\,,
\end{equation}
where $V(x)$ is a white-noise potential, drawn from the probability distribution 
\begin{equation}\label{whitenoise}
P[V] = {1\over {\cal Z}}\, \exp -{1\over 2D}\int_0^L\,  V^2(x) \, dx\,.
\end{equation}
Here ${\cal Z}$ is a normalization constant and $D$ is the variance, i.e., $\langle V(x) V(y)\rangle = D\delta(x-y)$.  
The singular last term in (\ref{hamiltonian}) is the contact potential, describing the coupling of the disordered system to the perfect lead\footnote{The hamiltonian (\ref{hamiltonian}) can be obtained from the tight-binding hamiltonian used in \cite{ks1,ks2} in the usual manner, namely, by sending the bulk hopping amplitude $t$ to infinity and the lattice spacing $a$ to zero, such that $t a^2 = \frac{\hbar^2}{2m} $ is held fixed. In particular, the contact coupling $\lambda$ is obtained by demanding that $\frac{t'}{t} = \exp \frac{\lambda a}{2}\simeq 1$, where $t'$ is the hopping amplitude associated with the link which connects the disordered system and the lead.}. The lead is described, of course, by the free particle Schr\"odinger hamiltonian $H_{\rm lead} =  -\frac{1}{2} \partial_x^2 $ along the negative half-line $-\infty <x < 0$. By varying $\lambda$ we can control the coupling  of the disordered system to the lead. In particular, the limit $\lambda\rightarrow -\infty$ corresponds to closing the system and disconnecting it from the lead. 
Thus, the DOR should collapse onto the real energy axis and coincide with the density of states of the closed disordered system in this limit.

Resonances in our open system correspond to quasi-stationary solutions of the Schr\"odinger equation
\begin{equation}\label{quasistat}
H\psi(x,k) = \frac{k^2}{2} \psi(x,k)
\end{equation}
subjected to the outgoing wave boundary condition 
\begin{equation}\label{outgoing}
\psi(x,k)  = e^{-ikx}\,,\quad\quad {\mathrm Re}\,k > 0\,, {\mathrm Im}\,k <0
\end{equation}
in the lead $x<0$ \cite{LL,BPZ,hatano,hf,ks1,ks2}.  The condition ${\mathrm Re}\,k > 0$ in (\ref{outgoing}) is imposed because the outgoing wave propagates to the left in the lead, and  $ {\mathrm Im}\,k <0$ is imposed since the modulus of the resonance wave function has to grow into the lead. (Anti-resonances in our system, on the other hand, correspond to values of $k$ lying in the third quadrant of the complex $k$ plane.) \cite{hatano,hf}.

In addition to resonances, the open system has also true bound states at negative energies, which cannot leak into the lead. They correspond to solutions of (\ref{quasistat}) subjected to the outgoing wave boundary condition (\ref{outgoing}) in the lead, at values $k$ which lie on the positive imaginary axis (their wave functions therefore decay into the lead). 

Motivated by \cite{halpern,id2}, it is useful at this point to introduce the logarithmic derivative
\begin{equation}\label{logder}
f(x,k)  = {\partial_x\psi (x,k)\over \psi(x,k)}\,.
\end{equation}
It follows from (\ref{hamiltonian}) and (\ref{quasistat}) that inside the system $f(x,k)$ obeys the Riccati equation 
\begin{equation}\label{Riccati}
\partial_x f + f^2 = 2V(x) -k^2\,,\quad\quad 0<x<L\,,
\end{equation}
where $f(x,k)$ is subjected to the initial condition $f(L,k) = -\infty$, independently of $k$. This is due to the Dirichlet boundary condition\footnote{$\psi(x,k)$ either vanishes at $x=L$ through positive values and nonvanishing negative slope, or through negative values  and positive slope.} set upon $\psi(L,k)$, since the 
system is closed at $x=L$. In practice, we shall impose the initial condition 
\begin{equation}\label{icL}
f(L,k) = -c
\end{equation}
for some real constant $c$, and in the end take the limit $c\rightarrow\infty$.

Evidently, in the lead, $f(x,k)$ is constant for any solution of (\ref{quasistat}) that is subjected to the outgoing wave boundary condition (\ref{outgoing}). In particular, for such a solution, 
\begin{equation}\label{0-}
f(0-,k)  = -i k\,. 
\end{equation}
Therefore, due to the $\delta(x)$ term in (\ref{hamiltonian}),  we know that the logarithmic derivative has to jump to 
\begin{equation}\label{0+} 
f(0+,k)  = -i k - \lambda
\end{equation}
right across the contact, on the system's side. We see that in the limit $\lambda\rightarrow -\infty$, we are effectively imposing a Dirichlet boundary condition on $\psi(0+,k)$, which means closing the disordered system and disconnecting it from the lead, in accordance with the comment made above Eq.(\ref{quasistat}).

We now have all the ingredients required to solve for resonant states in our system. For a given instance of the random potential $V(x)$ drawn from (\ref{whitenoise}), and for a given complex wave-number $k$, we integrate the Riccati equation (\ref{Riccati}), subjected to the initial condition (\ref{icL})  {\em backwards} in $x$, all the way down to $x=0+$. Note that due to this integration backwards, $f(x,k)$ can only depend on values of $V(y)$ with $y>x$. More precisely, it is straightforward to deduce from (\ref{Riccati}) that 
\begin{equation}\label{functionalder}
\Gamma (x,y) \equiv {\delta f(x,k)\over \delta V(y)} =  -2 \theta(y-x) \,\exp -2\int_x^y\, f(z,k)\,dz\,.
\end{equation}
(We shall make use of this result in the next section.)  Once we have $f(0+,k)$, we compute 
\begin{equation}\label{determinant}
F(k;[V])  = f(0+,k) + ik + \lambda\equiv {\partial_x\psi(0+,k) + (ik + \lambda)\psi(0,k)\over \psi(0,k)}\,, 
\end{equation}
whose functional dependence on $V(x)$ is indicated explicitly.
Thus, from (\ref{0+}), 
\begin{equation}\label{F0}
F(k;[V]) =0
\end{equation}
whenever we have a resonance at $k$ (in which case $k$ lies in the fourth quadrant, off the coordinate axes), an anti-resonance ($k\in$ third quadrant, off axes), or a bound 
state at negative energy (in which case $k = i|k|$ is pure positive imaginary).  

Note that $F(k;[V])$ has also poles, which occur whenever\footnote{As indicated explicitly in (\ref{determinant}),  $F(k;[V])$ is the ratio of two functions. These are analytic functions of $k$. Therefore, poles can arise only as zeros of the denominator.}$\psi(0,k) = 0$, i.e., whenever $\psi(x,k)$ obeys the Dirichlet boundary condition at $x=0$, in addition to the boundary condition $\partial_x\psi(L,k) + c\psi(L,k) = 0$ it is already subjected to from (\ref{icL}). Thus, these zeros correspond to eigenstates of the Hermitian system (closed at $x=0$), whose energy eigenvalues $E = \frac{1}{2}k^2$ can be either positive (with $k$ real), or negative (with $k$ pure imaginary, on the positive imaginary axis). It is the positive-energy eigenstates of the Hermitian disordered system which become resonances upon connecting it to the lead at $x=0$. Its negative energy eigenstates, while suffering slight distortions and energy shifts due to the contact potential, cannot leak out into the lead, and remain bound states of the open system, as was mentioned above.

To summarize, poles of $F(k;[V])$ lie exclusively on the real and imaginary axes of the complex $k$-plane.  They correspond to eigenstates of the closed Hermitian system. Zeros of $F(k;[V])$ correspond either to resonances, anti-resonances, or to genuine negative energy bound states of the open system. We thus conclude that $F(k;[V])$ is essentially the ratio of two spectral determinants: the spectral determinant of the non-Hermitian hamiltonian, which gives rise to resonances, divided by the spectral determinant of the Hermitian problem\cite{coleman}.

The quantity $F(k;[V])$ is a holomorphic function of $k$, which follows immediately from the holomorphy of wave function $\psi(x,k)$.  Counting resonances in our system is thus equivalent to counting the zeros of  $F(k;[V])$ in the appropriate region in the complex $k$-plane, namely, the fourth quadrant (and off the coordinate axes). To this end, we follow \cite{FZ,sommers}. 
Let $k_0$ be one of the zeros of  $F(k;[V])$. For simplicity, we shall take it to be a simple zero\footnote{We lose nothing by restricting to simple zeros. Indeed, if $k_0$ is a multiple zero of order $n$,  then $F(k;[V])\simeq \frac{F^{(n)}(k_0)}{n!}(k-k_0)^n$, and as should be clear from the discussion below, it will be simply counted $n$ times.}. Thus, in the vicinity of $k_0$, $F(k;[V])\simeq F'(k_0)(k-k_0)$, and therefore\footnote{Here  $\delta^{(2)} (k) = \delta({\mathrm Re}k)\,\delta({\mathrm Im}k)$.} 
\begin{equation}\label{k0}{1\over\pi}{\partial^2\over \partial k^*\partial k} \log \left |F(k;[V])\right|^2 = \delta^{(2)} (k-k_0)\,, 
\end{equation}
where we used the identity $ {\partial^2\over \partial k^*\partial k} \log |k|^2 = \pi\delta^{(2)} (k)\,.$ Similarly, a pole in  $F(k;[V])$ at $k=k_0$ (which must lie either on the real or the imaginary axis), would result in (\ref{k0}) with a negative sign on the RHS.

Off the coordinate axes, $F(k;[V])$ has only zeros, not poles. Thus, $\partial_{k^*}\partial_k\log \left |F(k;[V])\right|^2$ is a sum of strictly positive terms of the form (\ref{k0}). Let us denote the set of all these zeros of (\ref{determinant}) by ${\cal S}$. Therefore, we obtain the density of these zeros as  
\begin{eqnarray}\label{dor}
\rho(k,k^*, [V]) &\equiv& \sum_{k_n\in {\cal S}} \delta\left({\mathrm Re}(k-k_n)\right) \delta\left({\mathrm  Im}(k-k_n)\right)\nonumber\\{}\nonumber\\ 
& = & {1\over\pi}{\partial^2\over \partial k^*\partial k} \log \left |F(k;[V])\right|^2\,, 
\end{eqnarray}
where $k$ is kept off the coordinate axes. In the fourth quadrant ${\mathrm Re} k >0, {\mathrm Im} k <0$, the distribution $\rho(k,k^*, [V])$ coincides with the DOR of our system, for the given instance $V(x)$ of the disorder potential.

\section{The Average Density of Resonances}
We would like now to average (\ref{dor}) over the disorder (\ref{whitenoise}), in order to obtain the average DOR 
\begin{equation}\label{averagedor1}
\rho(k,k^*) =  {1\over\pi}{\partial^2\over \partial k^*\partial k} \,\Big\langle \log \left |F(k;[V])\right|^2\Big\rangle\,, \quad\quad {\mathrm Re}\,k > 0\,, {\mathrm Im}\,k <0
\end{equation}
of the disordered system. To this end we need the probability distribution function (pdf) of the random variable 
$f(0+,k)$, which enters (\ref{determinant}). In other words, we need the pdf for the event that $f(0+,k) = u + i v$, namely, 
\begin{equation}\label{puv}
P(u,v) = \Big\langle \delta\left({\mathrm Re}f(0+,k) - u\right) \delta\left({\mathrm  Im}f(0+,k) -v\right)\Big\rangle\,.
\end{equation}
(In order to avoid notational cluttering, we have suppressed the explicit dependence of this distribution on $k, k^*$.) Then, from (\ref{determinant}), (\ref{averagedor1}) and (\ref{puv}), we obtain the integral representation 
\begin{equation}\label{averagedor}
\rho(k,k^*) =  {1\over\pi}{\partial^2\over \partial k^*\partial k} \,\int_{-\infty}^{\infty}\,du\,dv\, P(u,v)\,\log \Big |u + iv + i k + \lambda\Big |^2
\end{equation}
for the average DOR, in terms of the pdf $P(u,v)$ for $f(0+,k)$.  Eq. (\ref{averagedor}) is our main result. 

Thus, we need to determine $P(u,v)$. This we do by extending the well-knwon treatment of the disordered {\em closed} system in \cite{halpern} (which is neatly explained and summarized in \cite{id2}), to the case of open systems. Thus, we introduce the pdf 
\begin{equation}\label{pdfx}
P(u,v;x)  = \Big\langle \delta\left({\mathrm Re}f(x,k) - u\right) \delta\left({\mathrm  Im}f(x,k) -v\right)\Big\rangle
\end{equation}
for the event that $f(x,k) = u + i v$. We observe that the Riccati equation (\ref{Riccati}) for $f(x,k)$ can be interpreted as a Langevin process in the complex plane, with the coordinate $x$ playing the role of ``time". By standard methods of the theory of stochastic processes (which we do not  belabor here),  we can derive the Fokker-Planck equation 
\begin{equation}\label{FP}
\partial_x P  = - 2D \partial_u^2  P + \partial_u [(u^2 -v^2 + 2 E_r)P] + \partial_v [(2 uv + 2 E_i)P]
\end{equation}
for $P(u,v;x) $. Here $E_r$ and $E_i$ are, respectively, the real and imaginary parts of the complex energy  $E = \frac{k^2}{2}$. We should perhaps just mention that in deriving (\ref{FP}) we used the relations $${\mathrm Re} {\delta f(x,k)\over \delta V(x)} = -2\theta(0) = -1\quad\quad {\rm and}\quad\quad  {\mathrm Im} {\delta f(x,k)\over \delta V(x)} = 0\,,$$ which follow from (\ref{functionalder}).

A couple of  remarks on (\ref{FP}) are in order at this point: First of all, note the diffusion term in (\ref{FP}). The minus sign in front of $\partial_u^2 P$ arises because ``time" runs backwards. The diffusion constant is of course positive. Second, note that there is no diffusion term in the $v$ direction. The reason for this is that $V(x)$ is strictly real. If we allow also a white noise imaginary part for $V(x)$ with variance $\varepsilon$, which is uncorrelated with its real part, we will have a diffusion term $-2\varepsilon \partial_v^2$ in the $v$ direction as well. In fact, it can be shown that the resulting Fokker-Planck equation in this case is related, via a similarity transformation (i.e., by a gauge transformation with an imaginary vector potential) to a Schr\"odinger equation, in euclidean time, for a particle with anisotropic mass (assuming $D\neq\varepsilon$), with an inhomogeneous  magnetic field $B\propto v$ perpendicular to the $uv$ plane, and a quartic potential.

Let us now resume our discussion of $P(u,v)$. By integrating (\ref{FP}) backwards, from the initial point $x=L$, where (\ref{icL}) implies the ($k$ independent) initial condition
\begin{equation}\label{icFP}
P(u,v;L)  = \delta(u+c)\,\delta(v)\,,
\end{equation}
all the way down to $x=0$, we can determine the desired pdf $P(u,v;0) \equiv P(u,v)$.  In the limit of very long system $L\rightarrow\infty$, the pdf at $x=0$, $P(u,v)$, which goes into (\ref{averagedor}), should be governed by the simpler stationary Fokker-Planck equation 
\begin{equation}\label{statFP}
- 2D \partial_u^2  P + \partial_u [(u^2 -v^2 + 2 E_r)P] + \partial_v [(2 uv + 2 E_i)P] = 0\,.
\end{equation}
It is straightforward to check,  directly from (\ref{statFP}), that the leading asymptotic behavior of $P(u,v)$ as $r^2 = u^2 + v^2 \rightarrow\infty $ is 
\begin{equation}\label{asymptotic}
P(u,v) = {1\over r^4} + {\cal O}\left({1\over r^5}\right)
\end{equation}
(up to overall normalization).  Thus, $P(u,v)$ decays faster than the minimal rate required by normalizability, and in a manner which is independent of $k$. In fact, more careful analysis shows that $k$ (or energy) dependence appears for the first time only in the $r^{-6}$ term.

Note that in the case of a closed disordered system, discussed in \cite{halpern,id2}, with Dirichlet boundary conditions at both $x=0$ and $x=L$, $f(x,k)$ is purely real (all eigenstates $\psi(x,k)$ are real, up to an overall phase). Thus, all reference to $v$, the value of ${\mathrm Im} f(x,k)$, disappears from (\ref{FP}) and 
(\ref{statFP}). In particular, the stationary Fokker-Planck equation in this case is simply  $$- 2D {d^2 \over d u^2}   P + {d\over d u} [(u^2 + 2 E)P] = 0\,,$$ with real energy $E$, and which can be solved explicitly, of course, by elementary methods.  At the moment, we are making an effort to derive a certain integral representation for $P(u,v)$, the solution of (\ref{statFP}), which we shall report on in a subsequent publication. 

\vspace{1cm}
{\bf acknowledgements}~ I wish to thank Boris Shapiro for many valuable discussion on resonances in disordered systems. 
This work was supported in part by the Israel Science Foundation (ISF). 


\newpage
\begin{thebibliography}{99}

\bibitem{LL}   
L. D. Landau and E. M. Lifshitz, {\em Quantum Mechanics: Non-Relativistic Theory, 
Course of theoretical physics , vol. 3} (Pergamon, Oxford, 1977). 

\bibitem{BPZ}  A.I. Baz, A. M. Perelomov and I.B. Zel'dovich, {\em Scattering, Reactions and Decay 
in Nonrelativistic Quantum Mechanics} (Israel Program for ScientiÞc Translations,  Jerusalem, 1969). 

\bibitem{hatano} N. Hatano, K. Sasada, H. Nakamura and T. Petrosky, Prog. Theor. Phys  {\bf 119} (2008), 187. 


\bibitem{hf} N. Hatano, T. Kawamoto and J.  Feinberg, {\em Probabilistic Interpretation of Resonant States }, these proceedings. 


\bibitem{wied} J. J. M. Verbaarschot, H. A. Weidenm\"uller and  M. R. Zirnbauer, Phys. Rep. {\bf 129} (1985), 367. 

\bibitem{rotter} J. Okolowicz, M. Ploszajczak and I. Rotter, Phys. Rep. {\bf 374} (2003), 271. \\ 
I. Rotter, Rep. Prog. Phys. {\bf 54} (1991), 635; ~ J. Phys. A{\bf 42} (2009), 153001. 

\bibitem{fyod-som} Y.V. Fyodorov and H.- J. Sommers, J. Math. Phys {\bf 38} (1997), 1918; ~ J.Phys.A {\bf 36} (2003), 3303.  (Special Issue on 
ÒRandom Matrix theoryÓ.)

\bibitem{datta} S. Datta, {\em Electronic Transport in Mesoscopic Systems} (Cambridge 
University Press, Cambridge, 1995.)

\bibitem{kottos} T. Kottos, J. Phys. A{\bf 38} (2005), 10761.  (Special issue on ``Trends in 
Quantum Chaotic Scattering".)  

\bibitem{ks1} H. Kunz and B. Shapiro, J. Phys. A{\bf 39} (2006), 10155. 

\bibitem{ks2} H. Kunz and B. Shapiro, Phys Rev. B{\bf 77} (2008), 054203.

\bibitem{terraneo}  M. Terraneo and I. Guarneri, Eur. Phys. J. B{\bf 18} (2000), 303. 

\bibitem{texier}  C. Texier and A. Comtet, Phys. Rev. Lett. {\bf 82} (1999), 4220. 

\bibitem{pinheiro}  F. A. Pinheiro, M. Rusek, A. Orlowski, and B. A. van Tiggelen,  Phys. Rev. E{\bf 69} (2004), 026605. 

\bibitem{tit-fyod}  M. Titov and Y. V. Fyodorov, Phys. Rev. B{\bf 61} (2000), R2444. 

\bibitem{weiss}  M. Weiss, J. A. Mendez-Bermudez, and T. Kottos, Phys. Rev. B{\bf 73} (2006), 045103. 

\bibitem{halpern} B. I. Halperin, Phys. Rev {\bf 139} (1965), A104. 

\bibitem{id2} C. Itzykson and J.-M. Drouffe, {\em Statistical Field Theory} (Cambridge U.P., Cambridge 1989), Vol. 2, Chapter 10.1.  

\bibitem{coleman} S. Coleman, {\em Aspects of Symmetry} (Cambridge U.P., Cambridge, 1985), Chapter 6, Appendix 1. 

\bibitem{FZ} J. Feinberg and A. Zee, Nucl. Phys. {\bf 501} (1997), 643; ~{\em ibid.} {\bf B504} (1997), 579. 

\bibitem{sommers} F. Haake, F. Izrailev, N. Lehmann, D. Saher and H.-J. Sommers, Z. Phys. {\bf B88} (1992),  359; H.-J. 
Sommers, A. Crisanti, H. Sompolinski and Y. Stein, Phys. Rev. Lett. {\bf 60} (1988), 1895. 


\end{thebibliography}
\end{document}